# Plan for Building a 1000 Person Martian Colony

**Vincenzo Donofrio**
Michigan State University, East Lansing, MI
donofr19@msu.edu

**Meghan Kirk**
Michigan State University, East Lansing, MI
kirkmegh@msu.edu

**ABSTRACT**

The implementation of a feasible colony on Mars has been discussed and analyzed for some time. It is noted that Mars offers all necessary in-situ resources to allow for self-sustainability, with the most efficient techniques and systems to produce water, food, metals, plastics, and other materials to be discussed. However, the solution to the economic viability of colonizing Mars is still in question. An initial mission is expected to cost upwards of hundreds of billions of dollars, which is generally the average to high end price settled on in most discussions, so it is necessary to generate an economic plan that requires a small fraction of this cost, as well as detailing exactly how quickly the colony can return on this investment and begin to profit. The focus of most topics will be split upon three general points in the population: one thousand, hundreds of thousands, and millions. Important decisions involving government and society will be discussed, as well as the need for an aesthetically pleasing colony. Additionally, our given scenario assumes that prior missions lacking human presence on the planet will take place, which will allow for the set-up of nuclear generators by way of robot and the continued robotic research that has taken place for decades now. An in-depth discussion of the distribution of this source as well as any other form of power will still be necessary. We also can assume that irrigation water and fertilizers have been found on Mars, with the water assumption being highly reasonable at this point considering the planet is almost certain to contain reservoirs of ice and perhaps liquid water and the fertilizer narrative leaves us with a nice bonus. Furthermore, the expected costs of shipping goods from Earth to Mars and then from Mars to Earth will be $500/kg and $200/kg, respectively.

**INTRODUCTION**

Before we dive into the building of a successful colony, we should outline why Mars is the best available option in the entirety of the solar system. More specifically, the discussion should be why Mars is factorially better than the second-best option humans realistically have: The Moon. The advantages the Moon possesses over Mars are almost non-existent barring a few. Earth's Moon quite clearly has the distance from Earth as well as the required velocity to escape it in its favor, with maybe a couple other advantages that serve no real purpose. While the distance notion really has no counter to it, the issue with escape velocity does. While it is easier to escape the moon, that comes at a huge cost. For a massive body to have this advantage, it means that its gravity well is generally more lenient, which is the case for the moon. With the moon being protected by literally no atmosphere whatsoever and the gravity on the surface coming in at an excruciating 1/6 of Earth's compared to Mars' manageable atmosphere and 1/3 of Earth's gravity, the advantage of an easier time leaving isn't worth the negatives. In fact, just a glimpse at the moon using the naked-eye will leave its holder fascinated by the massive craters that likens earth's moon to an "asteroidal dartboard", so-to-speak. The plethora of advantages Mars has over the Moon only rests this case. The combination of a workable atmosphere comprised heavily of carbon dioxide as well as other useful gases and manageable gravity makes Mars the most comparable candidate to Earth in the entire solar system. The necessary resources a colony will need to thrive are all available as well, with the moon being comprised of very little in comparison. The biggest advantage will be its water; while the Moon has recently been discovered to contain water, its amount hails in no comparison to the amount on Mars. Perhaps this discussion should be geared towards a comparison between an exoplanet like Kepler 438b and Mars, but its category alone should indicate why there isn't one to be had as of today.

**PRIOR TO HUMAN ARRIVAL**

Another conversation to be had before a human steps foot on the planet is regarding what will be brought. We will discuss the necessities and their respective quantities, with the approximate costs of the totals exemplified in the more relevant economic section.

If we want to really lower the cost, then it is important to choose the correct items and how much of each will be brought. It is crucial to think about what will essentially be available to make right away on Mars and then what will not. Also, the crew will need to factor in the probable eight-month journey they will spend traveling through space, which will be discussed first. The colonists will need a considerable amount of food and water to survive the flight. While what types or kinds of foods and beverages aren't of upmost importance, it can be assumed that a resounding amount will in fact just be water and various healthy greens for the small, fit crew. It is best to simply base how much should be brought around the recommended amount these sized people should consume for eight months. This means that for an eight-month period, around 800,000 kilograms of food and water should be brought to meet those requirements. An additional 250,000 kilograms of various plants and foods should be brought to give the crew enough food for a few months until the first imports arrive.

The crew will also have to bring resources that will serve great purpose on arrival and sometime after, too. While they should and will not bring any metals, plastics, glass, or fabrics, these can be made simply using in-situ resources on Mars, the crew should consider bringing shelter, spacesuits, and advanced electronics. This will also contain the nuclear generators as well as any robots that will provide great assistance to the humans, but as previously stated these will be brought prior to this mission with the amount (approximately 100,000 kilograms) still being factored into the total cost.

Although the colony will initially be placed underground with the colonists being able to walk freely in their tubes, they will still want spacesuits. Considering that there will be no available shops that will allow the crew to buy their suits, the only option for the initials will be to bring them. There are two types of spacesuits to choose from as well: one that emphasizes comfort and one that specifies safety. In technical terms, these two options are referred to as elastic and pneumatic suits, respectively. While the design and obvious comfort of the elastic suit will appeal to most, the wise decision is to go pneumatic. The use for a spacesuit on Mars is to protect a colonist from the cold, allow for breathable air, while also offering sufficient protection when traversing the surface prior to terraforming the planet. And with current suit technology, pneumatic is the option that more effectively meets these needs. A pneumatic spacesuit, specifically a Mars Suit[1], allows for an overall warmer experience, especially for the lower half of the body, as well as superior protection. Elastic suits will tear easier than pneumatic which practically will have no problems in that regard. In terms of pricing, elastic does offer the much lower price (a factor of 10 less), and as previously mentioned is more comfortable, but these advantages do not outweigh those that the pneumatic suit offers. If the crew wants comfort, they will make their own synthetic clothing upon arrival, in addition to the sweats they will undoubtedly bring. The total price of the suits will be only $32.5 million, anyway, with the advantages they give valued at priceless.

The crew will also have to bring their own shelter. The other option that is available is to bring enough shelter for maybe half the colonists, or simply not bring any at all and rely on instantaneously manufacturing tubes upon arrival, but to lower risk and boost safety, there will be enough shelter for everyone brought from Earth. Future structures will be made using in-situ resources, but for now the various greenhouse, living, and research tubes, as well as any helpful manufacturing items, will come in at a mass of 445,000 kilograms. Any advanced electronics that will be used will weigh in the range of the tens to hundreds of thousands, as well.

It is important to consider leaving some room for error, as well as understanding the idea that full self-sustainability will not be possible until the population reaches the hundreds of thousands, at least 25 to 50 years later. This means that the colony will still need to rely on and map out imports from Earth for at least a few decades, with the loads that comprise them decreasing at a likely exponential rate. Water can be exempt from this list, as it is straightforward to produce on Mars. The materials required for shelter and structures will also not need to be imported, say, a few years after initial settlement as the advancement of 3D printing technology will offer great aid to the colony. Food (mainly plant) imports will most likely be a mainstay for a considerable amount of time because the experimental procedures and testing on Mars that has to be done using terrestrial plants will probably offer the biggest challenge. Synthetic biological techniques are currently in the works[2], but until this can be perfected for Mars, food imports will be required. Factoring in this hopeful success, we will deem these imports to gradually decrease, but still be a large part of the load. Suits can easily be made on Mars for future colonists, and the generators should be made using in-situ resources, as well. Advanced electronics may be a mainstay, too, but not if food will be.

If the following parameters are followed, then there will need to be around $500 billion to $1 trillion invested if we are trying to be as efficient while offering some risk as possible, with numbers easily reaching into the double-digit trillions if a safer approach with more funds is available. While these projections are very high, keep in mind they are the totals until self-sustainability. As the colony reaches this point, the cost will continually decrease until virtually nothing will to be spent on imports, anymore. In a later section, we will discuss how to deal with these costs, as well as how much the initial mission will be valued at. The initial mission and the future cargo shipments should be separated financially, as they both coincide with their respective periods of economic success that will see profits starting right away in the initial phase as well as beyond these later shipments. As stated, specifics will be discussed.

**LOCATION**

The ability to pick the desired location of a colony is something that is unique to Mars. All the major human explorations of the past didn't allow for the ability to really scope out the targeted terrain. This can even be said for the moon. Although there were no real intentions to start a colony on the Moon, the technology that was used to study it before arrival doesn't even come close to the technology that has studied Mars, specifically over the last couple decades. This means that picking the perfect location for the colony is crucial and shouldn't be taken lightly.

There are many options that would allow for a successful colony to be built on Mars, but quite frankly only a few should seriously be considered. Factors that should be of highest value when discussing location are the availability of necessary resources, the condition of the terrain (i.e., not on top of volcano), and the opportunity for economic success that may come from tourism. Considering this, the two specified and prominent locations to be compared are the areas near Gale Crater and Valles Marineris (Mariner Valley).

Gale Crater would be a wise place to choose in terms of tourism. Imagine the attraction that would be centered around this massive 154-kilometer-wide crevice. Additionally, a colony placed near the edge will overlook the crater's enormous mountain, Aeolis Mons, which is located near its center and rises an incredible 5.5 kilometers, a sight incomparable on Earth. The depth of the crater may also allow for easier access to the possible huge water reservoirs beneath its surface when necessary. Gale Crater also offers a level of relatability per se to humans because it is thought to once be a lake many years ago[3], with the idea that it will be the first lake the humans engineer upon terraforming Mars. However, while this spot is a very reasonable landing place, Valles Marineris simply beats out Gale Crater in every way.

Valles Marineris is quite comparable to Gale Crater in that it also has great depth but reaches down to an even more extraordinary 8 kilometers in many areas. This will in turn allow for easier access to extradite water beneath the surface, especially considering that the valley is hypothesized to possess spring-like deposits[3], as opposed to the standard ice beneath it, bingo! Located near the equator, the valley will experience average highs of zero degrees Celsius. While not so much should be focused on temperature, location near the equator will maximize the sunlight available that will be used as solar power. As for another measurable, the valley is an astonishing 4,000 kilometers long which is a staggering five times longer than earth's Grand Canyon, its closest comparison, and as long as the United States itself! The Grand Canyon is one of the most attractive tourist destinations on earth, which only means that the Valles Marineris could be one of the most attractive sites in the solar system. Exact location should be pinpointed on either of its shorter sides, in the need to go around it, with the exact side being towards the east near the region Eos Chasma, which is thought to be resourceful.

# UPON ARRIVAL–VARIOUS TECHNICAL SYSTEMS

After the crew endures the long trip and eventually arrives on Mars, it will be important to start off on the right foot. This means being able to construct the various systems and conditions in place to raise the chances of survival and efficiency. Immediately, the crew will need to set up shelter, begin creating their systems to store energy, undergo the retrieval of water and production of food, and start to research and explore with the aim to quickly produce all remaining materials.

**Shelter and 3D Printing**
The crew can use the provided robots to dig out holes to place and assemble the inflatable tubes and greenhouses along the valley, a relatively quick and easy process. Some things to consider, however, are where and how far apart the shelter should be placed to avoid any shadowing problems with the eventually implemented solar panels. The crew can simply spread out the tubes, solving this problem. Using the reference provided by MarsSociety[4], it is safe to map out an area of three million square feet; this will allow for the colony to be a square with 1,700 feet on a side with six city blocks of 300 feet which will allow for the colonists to easily walk between labs and shelters when needed. Again, referring to the reference, the tubes should be 16 feet in diameter to limit costs of shipping. They will be pressurized at eight PSI to provide a safety factor of four against the pressure of two PSI that is evident at about eight feet underground on Mars. There will be a transparent strip along the top, with aluminized mylar reflectors being used to reflect sunlight into the tube.

3D printing has opened a range of possibilities that will improve a colony on Mars. The colonists could easily print out any object or tool quickly, instead of relying on imports. More importantly, this concept will thoroughly be used in the construction of above ground structures.

As the colony expands, massive domes, perhaps made of ice[5], will be made. But the cheapest and most efficient option will be to make use of this ever so interesting concept of 3D printing. A competition[6] held by NASA that looked to find the best concept for an above ground home ended with many exciting designs chosen as finalists. The winning group, Team Zopherus, designed a moving printer that would deploy rovers to collect local resources and produce the intended habitat. The use of technology like this will pay dividends for the colony.

**Energy/Power**

Picking the correct sources and then efficiently distributing them to power the colonists' various systems will be key. On Mars, the two main sources of energy will be through solar and nuclear power.

Solar power will be extremely useful but shouldn't be relied on. Considering the devastating dust storms that could engulf the colony for months at a time, only a minimal amount of power that the colony uses should be solar. Additionally, dust storm or not, the amount of dust that will settle on the panels during a normal day will be a hassle to clean off and will lower the amount of sunlight received. We will also have to factor that solar irradiance on Mars is 595 watts per square meter, nearly a factor of three less than Earth's, which comes as a result of Mars being further away from the Sun. So, solar power should exclusively be used in processes like water electrolysis as well as heating the living spaces and greenhouses.

Nuclear power will be relied on heavily by the crew, which will require the instillation of a plant above ground. Both nuclear fusion and fission reactions will be used, with the isotope deuterium, commonly found on Mars, being applied to both (deuterium will be discussed more later). Nuclear fission is and will be the preferred source, as it the most compact and least complicated to deploy. The crew can specifically use NASA's Kilopower reactor[7] for fission and one bigger generator for the rest of the needs. For a colony of 1,000 people, 2MWe is needed[4]. This means that 100 Kilopower reactors (10 kWe) can combine output with the one MWe generator, which will be a cheaper option to bring when considering import taxes. The total mass for the previously stated scenario will be around 40,000 kilograms, half that of a scenario that involves twenty 100 kWe generators each weighing 4,000 kilograms: a total of 80,000 kilograms.

Nuclear power will be able to keep the colony warm and running during dust storms, as the power plant will produce a thermal output of 27 million btu per hour (British thermal unit), compared to the nearly 20 million btu per hour the greenhouses will lose[4].

**Water**

For our scenario, it is a given that irrigation water has been found. As mentioned before, this is a wise assumption, but an overview of the process to retrieve more water is necessary. The most efficient way would be to assign the robots to extract the ice water underground by way of mining and melting of ice.

The availability of water also allows for some crucial gases to be obtained, too. Water provides access to hydrogen, though the electrolysis of water, which is an important component of fuel (liquid hydrogen). Oxygen is also a product of this process, the remaining component in the liquid hydrogen fuel. Additionally, hydrogen can be used to make certain plastics, with oxygen of course allowing the colonists to breathe.

**Food**
The production of food may be the most difficult process to complete for the colonists. Realistically, food (plants) will have to be imported for a while until the colony understands how Earthling plants grow in Martian soil. For food to grow using Martian soil in the greenhouses, the regolith that contaminates it can be cleaned using the water the colonists produce. On Mars, the practice of plant synthetic biology can be further researched[2], with the goal of engineering plants to adapt specifically to the Martian environment and end the dependence on shipping food from Earth. Of course, the plants the crew grow will provide them with the oxygen they need in the tubes.

**Glass**
The production of glass, and specifically fiberglass early on, will be useful for the crew. About 40% of the weight of the soil on Mars is made up of silicon dioxide: the basic component in glassmaking. Unfortunately, Martian soil is also heavily made up of iron oxide as previously mentioned, which will leave the colonists will a lower quality of glass. However, this isn't really a problem for the colonists who will want to opt for the most efficient processes, and the need to remove the iron oxide through a tough process isn't necessary. Therefore, the tinted silicon dioxide glass can be made using the same sand-melting techniques used on Earth.

As the colony grows, a good shout would be to build an above ground city and dome out of this glass, preferably with the iron oxide removed. This would be wise as something like glass will be more available and most likely easier to make than cement, but that is something that will be interesting to discover.

**Vehicles and Transportation**
As the colony grows and walking becomes inefficient and tiring for the crew, there will be a need for an upgrade in transportation. The first need will be in the form of a simple bicycle. Mars offers two materials that can make up the frame of the bike: steel and aluminum. Additionally, the seat and handlebars and what cannot be made of the plastic that the colonists can produce, as well. The wheels should be of required quality to traverse through the thick sand using technology already present on earth.

Eventually, as the colony expands into millions of people and inhabits a variety of regions across the globe, more sophisticated technology will be preferred. This is also where things will get a little interesting. But first, the obvious choice of transportation, as is on earth, will be by car. If we stick to the idea of using clean energy, then specifically we want electric cars. And on the rocky and bumpy surface Mars will provide inhabitants, a sort of jeep modification will be the type of car selected. While roads will eventually be placed, they are not of top priority for travel on a planet that will not have any vegetation or bodies of water to get in the way for a while, but simply just a whole bunch of red sand.

The interesting part is when the discussion of air travel on Mars is brought up. While travel by the earth's standard commercial airplane is quite bad for the environment, a less threatening option is by way of rocket hopper. More specifically, the most viable option that is available currently is Spacex's Starship[8]. The ability to travel around the planet in an hour will be vital for the expansion of the civilization on mars, but the big concern that lies with this method still resides in its ability to use a clean, cheap source of fuel. This is because on earth, the production of the hydrogen and oxygen sources used to make the required liquid oxygen fuel mainly are generated using non-renewable forms of energy. However, renewable sources like wind and solar power could be used in the process, but at much higher costs. If on Mars comes a breakthrough in this technology that lowers costs and in turn allows these renewable sources to be efficient and useful, then a rising form of travel that will completely alter the aerospace industry may just be perfected and explode.

A form of transportation that will be wise to look out for on the future of mars will be by way of passenger-carrying-drones. Mars will certainly be a place where the risks that are just too much for Earth will be tried instead. While not necessarily a viable option until the dust storms are settled, which means probably not until the planet is terraformed, the use of this technology will surely be explored. Although these could easily just turn out as glorified helicopters, a slick Martian or two may just unlock the keys to a new, unique way to travel.

**Communication**
Communication during the human settlement of mars will be key. The routes that need to be covered are communication between Mars and Mars, and Mars and Earth and locations like the asteroid belt.

The most important form of communication, especially early on, will obviously be between Martians. Ideally for the few who are still left with exploration jobs, but useful for everyone, will be an accessible way of communication that is at the push of button and attached to one's spacesuit. To allow the radio waves to be transmitted between Martians, satellites will need to be implemented, something not entirely too difficult, as long as there is the means of transportation to get the required position in Mars' low orbit. This is something that can be implemented during prior, unmanned missions, with the cost being virtually nothing when added to the total amount of the entirety of these first missions to mars. Communication between Martians and Earthlings and those at other locations can be done through these satellites as well, specifically using laser transmission techniques to limit the transmission distance handicap. Another option for this long-distance communication would be to use smartphones or the like to simply message between planets. This will of course require an internet connection to be founded on mars, something that may impose greater difficulties than setting up satellites, but once invented will be used as another form of communication unfortunately again with a delay. Instantaneous communication between planets will perhaps be unearthed, but for now a slight delay will have to do.

**POLITICAL**

For the first 1,000 colonists, it may be quite difficult to establish a complete government solely on Mars, but basic laws and rules can be put in place. In a perfect scenario, many of the leading powers on earth will be united by this great cause, and with thorough negotiation, the first 1,000 colonists can be governed by this newfound united front, with possibly one key leader with a few other higher-ups to lead the other colonists.

Unfortunately, this will unlikely be the case, and either a private company (like SpaceX) with help from a nation's government and its space program (like the US and NASA) will lead the crew, or the crew will simply be guided entirely by said government. Using the former, highly likely scenario, there will be a local government on mars comprised of a democratically selected leader and board of officials as well as a strong presence from, in this scenario, the US government back on earth. This is seen in many other instances in colonialism throughout recent history, most notably, colonial America. Like our plan with the colonization of mars, colonial America was comprised of a few forms of local government, with also a great deal of interference from the British government. Of course, however, it is important to not impose the strict and unfair taxes and laws that eventually led to the breakaway of colonial America, something that absolutely cannot happen on mars in terms of unification. While seemingly this will be simply avoided because of the positive track record the democratic government of the US possesses in terms of the capability its citizens possess to express basic freedoms, there is still a need to specifically address these necessities. If Mars wants to ultimately end up as the massive economic success envisioned my many in the short-term future, mainly because of the unique industries and opportunities for great inventions and achievements, it will need to be allowed to freely express these avenues. This means that, unlike colonial America, the Martians will need to be given the ability to apply for and at least retain the rights to their creations.

The goal is to allow the chosen Earth government to oversee the colony, protect the rights and interests of its colonists, collect on the fairly priced export and import taxes, all with the confidence that Mars will quickly become economically, technically, and socially successful which will then lead to eventual self-governing metropolises. There will initially need to be a high-level of interest from Earth towards Mars, with a relatively steep decline that coincides with the expansion of the colony.

As the colony expands and soon is comprised of multiple cities with its population in the millions, this will coincide with changes in governmental structure. Like with many nations on Earth, the cities on Mars should consist of multiple types of initially healthy governments, based on the various needs for that society. The strongest and most successful ones will survive, and the ones with poor ideas will fail. While the objective isn't to deem cities as utter failures completely right away, as seen on earth, this will eventually happen as the planet becomes heavily populated. However, with the goal of keeping peace and unity between these cities that will arise from a culture that is built on togetherness and survivability, hopefully the planet will not experience much catastrophic failure and unhealthy competition that is prominent on earth. But hopefully a great level of support for those failing cities will be expressed, because at the end of the Martian day, it will be incredibly tough to succeed.

To ensure success and sustainability on Mars, these specific policies should be instilled or granted. First, the need for property rights will be crucial. As the colony expands, and people begin to look to structure their own housing, the last thing they will want is for some Martian pirates stealing it away. So, the Martian government will need to establish a basis for these property titles to avoid this problem. Additionally, the need for this process may even be called for sooner as undoubtedly businesses on Earth will want to stake an early claim in territory before the eventual skyrocketing of the price for the land.

As stated before, the rights of any invention created by a Martian should stay with the Martian. If the colony wants to succeed, they will have to retain the entire value of their products to hold up in the trading business. The Martians should also be able to apply for patents from Earthling marketplaces if they desire to grow their business independently, and not just be limited to only being able to bargain their inventions.

Efforts from nations that will try to disrupt the colony will surely be acted out, so it is important to have a politically diverse planet, with key regulations.

**SOCIETAL/CULTURAL**

Initially, the society will be made up of many values from earth. With only 1,000 on the planet, the colony will tend to be hierarchical, with one leader who has a board comprised of multiple officials. As the colony expands, there will inevitably be those that want to venture out and create new colonies, which will eventually lead to massive cities and metropolises. As the planet reaches into the millions and eventually gives birth to its third and fourth generations, perhaps these Martians will finally consider Mars their true home and prefer never to go back to Earth. This will raise the question of if and how the planet can hold values that are better than those of earth's society, with the need to avoid war and congressional conflicts. This can be done through thorough education, in both technical and cultural aspects in how to be and what it takes to be a Martian. Also, the forming of a society that can make better use of its time when it comes to their social, personal, and professional life, as well as the important need of increasing physical, mental, and emotional health is necessary.

Martian society will always be compared to the only other society in existence: Earthling society. While Mars will never be a utopia many claim it could be, it will most certainly be a place that will build off those values that Earthling society presents and build as close to a perfect society as possible, while in turn giving Earth something to compare itself to better its own society.

As the colony quickly progresses into the hundreds of thousands and multiple branched-off colonies begin to arise, the need to keep essential values that will instill the peace and creativity the planet was founded on will be crucial. While no need of formal education will be necessary yet, the implementation of certain groups and activities will be. On earth, this would be compared to one's religion, but of course the inhabitants of mars are allowed to keep their faiths. This expression will represent religious qualities in the sense that they will give many a sense of hope and unity, which will be much needed on a planet that is quickly expanding and distancing itself proximity wise. Perhaps a new form of social media will arise, maybe in the form of virtual reality, that will be able to link those who have grown apart and allow to keep those bonds that were formed out of pure love to lead a life of creativity and survivability, or simply to start a new, better one.

Maybe on Mars the need for in-person interactions, something ever so disappearing on earth, will arise, with many civilians longing to take a visit to their friends or even the other citizens of different cities, and spread the hope and creativity they've found on the planet. The expression of this strong, loving culture will most definitely be present in each respective city, with life almost being tribal like as each member knows how risk taking this life was and is, which will allow great support to everyone alike.

As the colony grows into the millions, and these colonies turn to major cities, the need for formal education will arise, as at this point there will be Martians who have been born on the planet. Martian school should consist of the technical and cultural necessities that allows for the survivability on mars. While this schooling for young Martians should of course still retain the basic mathematics, language (perhaps a new one will arise), and history of Mars and Earth alike, the education should be integrated in the students' lives. There should be great emphasis on what brought them to this wonderful planet, why being a Martian is so important. The technical studies should be in touch with the environment around them, teaching these kids how and why what they learn will be useful for them to perhaps one day colonize a planet in another solar system, and not just learn about what a concept is. As on earth, the need to educate youth properly to allow them to uphold values as well as knowledge that will let them use all their potential to positively impact society is a must, so hopefully Mars will display that more so than earth, which does a better job that most gives it credit for, anyway.

The day society on Mars breaks away from its core values will be a sad one, perhaps it will eventually occur, but if these steps are followed and kept in the back of the mind of future Martians, this won't happen.

An important mindset to keep will be the need to be as environmentally friendly as possible on Mars. This kind of goes without being said, but a fine job in this area will result in possibly the greatest difference the Martians will possess over the Earthlings. The focus can simply be split up into phases: pre- and post-terraform. In the early stages of the colony, the crew is essentially dependent on clean sources of energy: solar and nuclear. And if after a little while it is absolutely certain that Mars were dominant of life in the past and contains oil reserves beneath its surface, and the colony is tempted to use it as another source to power their stations, they must refuse at all costs. Or if they find reserves of a completely unknown but plentiful resource, and it is deemed to emit "dirty" gases, then another use that doesn't involve the burning of the resource will have to be found if the colonists want to implement it into their lives. The precedent of relying solely on clean sources of energy over a substantial period that will be set on Mars will prove to be of great importance and set a much-needed example for Earth.

In terms of post-terraforming, this same notion will have to be set. As the population disperses and people start to break off and inhabit all regions of the planet, they will be tempted to use their own energy as well. And if there are indeed oil reserves as mentioned before, rules will have to be placed that prevent the emission of the dirty gases that are released once these sources are burned even if they offer convenience for that group. With a fresh, completely clean earth-like atmosphere, there can and will not be any way that humans will make the same mistakes as they did with earth's atmosphere.

One of the most attractive narratives of Martian society that will sell many on Earth to make the trek to Mars will be its unique but promising job market. This also can be applied to the initial colonists. It is true that a labor shortage will quickly commence for the first inhabitants, as any remaining maintenance and exploring can more effectively be done by robots, with a handful number of humans needing to be hired to perform the odd repair on said robot. However, these colonists will be in great luck as there will be plenty of other jobs that will arise after this very early phase. For one, research on Martian plants, or the adaptation of Earthling plants, will need to be continued. Any heavy-hitting physical tasks can be performed autonomously, but for now humans should still be the primary conductors of the actual research. And not just plant-based research will be conducted, but plenty of other avenues all the way from terraforming to examining possibilities of life on Mars. Additionally, many colonists will be tasked with engineering and design of future above surface habitats and systems, or at least how to get the most out of them.

As for future colonists, as stated there will be a variety of options that will available. The most obvious should be the need for teachers. Unfortunately, on earth, teaching is not a very financially attractive profession, but serves a great purpose. However, on Mars, the job of a teacher will for some time be rewarding as well as important. As mentioned previously, the need to educate future Martian children will be of highest order to instill the values as well as the necessary skills to survive on the planet. But it won't be easy. A teacher on Mars will be someone who has lived generally at least 10 or so years on the planet and therefore has the necessary experience to convey crucial knowledge. Like with a good number of jobs available on Mars, a degree from a university, perhaps on Mars, will be necessary, and combined with the required experience will liken a Martian teacher more towards a college professor on Earth. So, the starting pay of the Martian teacher should be in the low six figures a year, especially considering that the school year will most likely be sanctioned at one full Martian year (687 Earth Days).

A plethora of other jobs prevalent on Mars can be discussed, too. On a planet where cryptocurrency will likely dominate, and reasonably should, the need for relevant data scientists and financial analysts will be prominent. An already popular industry on Earth, the "Airbnb industry" per se will be massive on Mars. Certainly, by this time, the production of unique Martian food will be perfected, which will only lead to a stream of restaurants on the planet. Of course, it can be inferred that many will be comprised of menus that focus on Earthling foods, too. The timekeeping industry will surely need to be reinvented, with a new line of expensive smartwatches to be released. Oh, and doctors, plenty of doctors, will need to be required, as well.

The colony will need to pay the salary of every job one way or another. Wherever the money comes from, whether if the job is with a private company or the Martian government is paying largely out of their own pocket with maybe some help from Earth, money will need to be generated through key industries.

**ECONOMIC**

Perhaps the most important discussion to be had when it comes to sending humans to Mars is the cost, and more specifically how to make a profit.

Realistically, the crew will need to be supported financially as, although these 1,000 people will be some of the most eager and willing to go, they are of course intelligent beings and will not want to go for absolutely free. Much of the work done on Mars can be compared to tasks done on Earth, with the obvious difference being the environment. Considering this rigorous environment, the crew on Mars will have to work in, their pay should be increased by a factor of 2 compared to the average rates on Earth. However, we will also consider that after a few months, labor will significantly decrease along with the pay. This would still leave the average colonist with a cool pay of around $20,000 per the first six earth months, which this rate dropping to $10,000 per earth month after that. This would leave any space agencies and partners to shell out another $200 million.

Given these initial investments that will cover one earth year worth of time, a goal to meet, say, one earth year after settling will be to break even or possibly even begin to profit on this whole ordeal, with heavy focus towards the latter. The good news is that Mars offers us quite a few minerals and ores that will be valuable trading options with earth. The bad news is that export costs will leave us with very little profit for most of these trading options, but not all.

While there is simply no real use to mine resources like gold or platinum on Mars, there is an incentive to mine deuterium, the heavy isotope of hydrogen. Deuterium is a very valuable source of fuel for nuclear reactors, while also having the capability to be used in the production of heavy water which in turn can be used as fuel for fission reactors. It is deemed valuable because of the incredible advantage in regard to abundance it has over the earth. On Mars, deuterium[9] occurs 833 times out of every million hydrogen atoms, while it only comprises 166 out of every million hydrogen atoms on earth giving mars a near five-to-one advantage. Additionally, considering that we will have majority of our power used in water electrolysis to run our various life support needs, we will be able to produce deuterium at virtually no extra cost. Deuterium is a by-product of electrolysis which would produce around one kilogram of deuterium for every six tons of water. Now, this is where we can make a case as to why the mining of these resources isn't necessarily worth it. A colony of 1,000 people would produce about 6,000 tons of water in the initial year, which means that only 1,000 kg of deuterium will be produced. Considering that pure deuterium is valued at approximately $10,000 per kilogram, this leaves us with a profit of $9.8 million after export costs. While it is a nice chunk of change for practically no additional work, the colony will only see an extreme value in deuterium when its numbers reach the hundreds of thousands. Until then, this unique isotope shouldn't be a trading priority for the crew, and even should seriously be considered to solely be used as an input to the nuclear reactors on the planet.

Sticking with the same sort of theme, though, a large portion of the debt can be made by using something simpler than a rare isotope: Martian sand. It is inevitable that collectors will want to get their hands on the very first samples of Martian sand and rocks brought back to Earth. The moon landing provides a precedent event of this scenario. Moon rocks[10] brought back from the various missions between 1969 and 1972 were valued at around $50,000 per gram with a grand total of around 400 kilograms of the stuff up for auction. One could only imagine that sand from the first planet colonized by humans in their attempt to make the species multiplanetary could fetch five or ten times that amount, but we'll stick with the precedent that was set by the moon samples, especially if our goal is to bring back say 1,000 kilograms which may decrease the value slightly. However, a high enough price will still be set, so a sensible amount to export will indeed be around 1,000 kilograms which would total to just under $50 billion in value after export costs. Let's say that the collectors on earth will not exhibit as much interest in our sand and only shell out $5 billion; well, that alone should cover the costs for this initial mission over the course of one year.

So, after one year we can continue to rely on sand and greedy collectors to make our easy money and supply future missions, right? Of course, the answer is no. It will be obvious to see that people can only handle so much sand and it can be easily assumed that after one auction of our 1,000 kilograms of it, the price will decrease exponentially and will soon cause the colony to lose money. Besides, there must be more sophisticated ways to make money on Mars than using its own sand, right?

The answer to this question will be… of course, yes! A huge industry will reside in the ingenuity of our own colonists. Using the example of nineteenth-century America, we can compare this frontier to our own. As seen with nineteenth-century America, our colony of 1,000 will quickly experience a labor shortage after all maintenance as well as trivial tasks will be in the hand of robots, with the only job the human will have will be to recalibrate said robot. In America, their shortage led to a flood of inventions in a society that didn't have nearly the same technological culture as us humans today and certainly the colony. So, will this explosion of ingenuity apply to the colony? Of course, it will! Unlike this period in America, Mars offers the crew a unique environment that will lead to some of the most extraordinary, cutting-edge technology in existence. Whatever field these inventions reside in, the licensing and patents for them will soon be worth perhaps trillions of dollars. Specifically, robotics and energy production will be amazing areas to profit from.

So, as it stands, we seem to have our financial needs covered, especially over the years as the colony, while continuing to exponentially expand, will also exponentially increase their ability to survive only using the in-situ resources Mars offers thus decreasing costs.

Well, as the colony reaches into the hundreds of thousands of people, say some 20 years after initial settlement, the ingenuity will level off. And as the explosion of ingenuity levels off, there will need to be more reliable industries and markets that will allow the colony to meet the costs of the remaining imports, needs, and salaries of its employees.

This calls for the need of some industries that could possibly work initially but are much better suited quite a few years down the road, with the most prominent being the mining of asteroids.

The mining of asteroids for high-grade metals will undoubtedly be a trillion-dollar industry, and it already has started on Earth, but the clear advantage would be to focus any company interested in this endeavor on Mars. Obviously, many of the mined asteroids will lay in the asteroid belt, a region that is significantly closer to

Mars than Earth and contains around 99 percent of the 800,000 or so known asteroids today. Additionally, the Near-Earth-Objects that make up the remaining one percent exclusively consist of asteroids that orbit closer to Mars, instead. The issue, as stated before, isn't if this industry will be successful, but rather how. Previously, critics would mention the risk of sending people to retrieve samples from these massive rocks, but fortunately humans will not have to go at all. The Japanese Hayabusa2[11] has already snagged a small sample from the asteroid Ryugu by firing a tantalum "bullet" as it neared. While the technology to mass mine multiple asteroids at large quantities may not be present today, surely there will be huge incentive to develop this technology on Mars. Nonetheless, this inevitable industry will undoubtedly lie in the hands of Martians, with Earthlings willing to pay a steep price for the riches that it will bring.

Another industry that will be very successful on Mars will be tourism. With the colony being positioned near one of the most fascinating tourist sites the solar system has to offer, a trip to Valles Marineris will trump all others on earth. While the tourism industry most likely will not explode until the permanent population reaches the tens or hundreds of thousands with the implementation of massive domes or structures specifically designed with hotels and resorts for tourists, it will be up there with mining if asteroids in terms of financial success for the colony and planet, generating billions if not trillions of dollars, as well. Of course, the engineering of technology that significantly reduces the time it takes to get to Mars will help the tourism industry succeed even more, but this will certainly be a huge market regardless.

So, for a financial round up, the colonists will require an initial input of $1.25 billion dollars. There will need to be around $400 million spent on transporting enough food and water for the colonists to survive the trip to Mars with another $125 million of additional food, a total of $32.5 million in suits, and another $400 million for the basic shelters, generators, robots, and advanced electronics.
Also, a good $100 million can be added to maintenance and miscellaneous costs. We cannot forget that about $200 million needs to be set aside to pay the salaries of the brave 1,000. Slightly inflating the total budget to around $1.5 billion, this number is definitely on the lower side, but still respectable and covers all of the necessary aspects the colony and colonists need. As stated previously, this first-year total with be met and exceeded with the sale of the in-demand deuterium ($9.8 million) and intriguing Martian-sand-collectibles ($5 billion). Additionally, as the amount of labor decreases, the rise of profitable inventions will take place which will allow those ingenuous colonists to make a fortune (billions and eventually trillions) and really allow us to see some serious return on investment.

As the colony expands, we need to pay the average citizen on Mars for their average job as well as a good possibility of $500 billion to $1 trillion in additional imports. When the population reaches a million, a good estimate is anywhere between half to three quarters will need to look for a job. While some jobs will pay better than others, the average salary is around $100,000 factoring the rigorous environment in. While say half works with a company that obviously pays their salary, another half will work in jobs funded by the government (the very trivial, busy work kind of jobs that will be needed and useful for newly arrived Martians). Each year with this number of Martians receiving their salary through the government, another $37.5 billion will needed to be generated. This will call for industries to take off in order to contribute to the pay for these jobs. Namely, the asteroid mining and tourism industries will explode coincidentally, generating billions and eventually trillions of dollars, as well. While not as explosive as before, the invention and creativity niche will still pull in massive numbers. A representation will be provided below to visualize all the financial factors and how each debt will be overcome over a respective length of time.

**AESTHETIC**

The need for the colony to look beautiful will be very important for the crew. During those times of struggle, the crew will need some psychological relief. Well, Mars presents all the psychological attributes one will need. Imagining looking out towards Valles Marineris and taking in the layered bands of rock which reveal millions of years of geological history. Additionally, the warm colors of the ground and sky give a lucid life. The colonists should also bring mementos from home and decorate their respective homes with them, which will provide a deeper level of beauty to the individual colonist, something of great importance, as well.

As the planet becomes fully terraformed, the beauty of the planet will really come to fruition. Imagine the nearby crater, previously dry for millions of years, soon to be completely filled with water and becoming a lake. Or maybe ponder how beautiful it will be to see the once empty lot of red land soon covered with rich forests and wildlife. The colony can even begin to bring back the once massive oceans the planet was once covered by. The dust storms can be traded in for a nice cool breeze that can relax anybody.

The real beauty can be found when the planet resembles the colonists' home of once upon a time. This accomplishment will offer great emotional appeal and finally allow the inhabitants to feel at home, a feeling that will be unmatched. The beauty that resides in the process to get to Mars will one day reveal itself on it.

# ACKNOWLEDGEMENT

The authors would like to acknowledge and give thanks to The Mars Society for the opportunity to participate in the Mars Colony Prize Competition.

# REFERENCES

1. Macdonald, Fiona. (2015). NASA's Released a Prototype of The Spacesuit Astronauts Will Wear on Mars. Accessed February 21, 2019, https://www.sciencealert.com/nasa-s-released-a-prototype-of-the-spacesuit-astronauts-will-wear-on-mars.
2. Llorente, Briardo. (2018). How to grow crops on Mars if we are to live on the red planet. Accessed February 10, 2019, https://theconversation.com/how-to-grow-crops-on-mars-if-we-are-to-live-on-the-red-planet-99943.
3. Lavars, Nick. (2016). Three great places to live on Mars. Accessed March 1, 2019, https://newatlas.com/great-places-to-live-mars/45654/.
4. Mole, Alan. (2019). Accessed January 29, 2019, http://marscolony.marssociety.org/MARSColonyi2.pdf.
5. Gillard, Eric. (2017). A New Home on Mars: NASA Langley's Icy Concept for Living on the Red Planet. Accessed March 25, 2019, https://www.nasa.gov/feature/langley/a-new-home-on-mars-nasa-langley-s-icy-concept-for-living-on-the-red-planet.
6. Harbaugh, Jennifer. (2018). Top Five Teams Win a Share of $100,000 in Virtual Modeling Stage of NASA's 3D-Printed Habitat Competition. Accessed March 20, 2019, https://www.nasa.gov/directorates/spacetech/centennial_challenges/3DPHab/five-teams-win-a-share-of-100000-in-virtual-modeling-stage.
7. Anderson, Gina. (2018). Space Technology Mission Directorate. Accessed March 1, 2019, https://www.nasa.gov/directorates/spacetech/kilopower.
8. Cavendish, Lee. (2019). Meet SpaceX's Starship Hopper. Accessed March 29, 2019, https://www.space.com/spacex-starship-hopper-elon-musk-explained.html.
9. Zubrin, Dr. Robert. *How to Live on Mars.* Accessed December 18, 2018.
10. Pearlman, Robert Z. (2011). NASA Busts Woman Selling $1.7M Moon Rock. Accessed March 9, 2019, https://www.space.com/11804-nasa-moon-rock-sting-apollo17.html.
11. Chang, Kenneth. (2019). Japan's Hayabusa2 Spacecraft Lands on Ryugu Asteroid. Accessed February 27, 2019, https://www.nytimes.com/2019/02/21/science/ryugu-asteroid-hayabusa2.html.